\documentstyle[psfig,preprint,eqsecnum,aps]{revtex}

\begin{document}

\title{Self-organization of (001) cubic crystal surfaces}

\author{ Laurent Proville} 
\address{Groupe de Physique des Solides, UMR 7588-CNRS\\
 Universit\'es Paris 7 $\&$ Paris 6, Tour 23,
 2 pl. Jussieu 75251, Paris Cedex 05, France}

\date{\today}

\maketitle

\begin{abstract}
Self-organization on crystal surface is studied as
a two dimensional spinodal decomposition in presence of a surface stress.
The  elastic Green function is calculated for a  $(001)$ cubic 
crystal surface taking into account the crystal anisotropy. \\
Numerical calculations show that the  phase separation 
is  driven by the interplay between domain boundary 
energy and long range elastic interactions.  
At late stage of the phase separation process, a steady state appears
with different nanometric patterns 
according to the surface coverage and the crystal elastic constants.

\end{abstract}

\pacs{64.70.Nd   , 68.43.Hn ,68.43.Jk ,81.40.Jj } 



\section{Introduction}

Self-organization (SO) on solid surface is an efficient mean for growing
nanostructures with regular sizes and spacings. 
The models proposed by
Marchenko \cite{Marchenko81} and Vanderbilt {\it et al.} \cite{Alerland88,Vanderbilt}
are the 
basis of the theoretical framework to understand the SO phenomenon. They enhanced
the interplay between the 
long range elastic interaction yielded by the underlying crystal
and the domain boundary energy.
Indeed, the former is minimum when two surface defaults are separated by a
distance as large as possible while the latter is minimum when only one compact domain 
appears onto the surface. So  when these two ingredients
are present (see below for experimental descriptions),
the surface ground state
structure should balance the aforementioned interactions.
The purpose of \cite{Marchenko81,Alerland88,Vanderbilt}
was to evaluate this structure in different cases.

Assuming  that two phases, called A and B, coexisting onto a crystal surface, 
have different intrinsic stresses, the authors of Refs. \cite{Marchenko81,Alerland88}
showed that a state which consists of stripes domains
occupied alternatively by A and B lowers the energy with 
a period selection. This period depends on the crystal
stress energy compared with the domain boundary energy and it varies exponentially
with the ratio of those two quantities. 
In order to perform an analytical surface elastic Green function calculation, 
Marchenko and Vanderbilt  both assumed either a crystal  anisotropy 
along one direction of the surface or the anisotropy of the intrinsic stresses.
Those works specially addressed the cases of  corrugated crystal surfaces and 
crystal surface reconstruction with broken symmetry
(see Ref.\cite{Au111,Si113} for recent relevant experimental analysis).

Comparing the stability of different periodic domains in an isotropic
2-dimensional dipolar model, Vanderbilt {\it et al.}  \cite{Vanderbilt} 
proved that  at
intermediate coverage, $\theta_0 > 0.28 $, 
the stripe structure is the optimal  candidate for the 2-dimensional system
ground state while at low coverage, $\theta_0 < 0.28$, 
the droplets structure is more stable.
As the dipole-dipole interaction is similar to the
elastic interaction, i.e., it decreases as the inverse of the distance to the power three,
these  Vanderbilt's results hold for
the solid surface SO. It shows that
the anisotropy of the intrinsic stress surface, assumed in Refs. \cite{Marchenko81,Alerland88}
is not essential in the SO process since the patterning occures even if 
the surface is isotropic with no symmetry breaking. 
The studies of Refs. \cite{Marchenko81,Alerland88,Vanderbilt} 
emphasize the main physical ingredients which drive SO, i.e. , 
the interplay between domain boundary energy and elastic interaction energy.

Recent analysis of chemisorbed mono-layers on (001) Copper surfaces, 
via Scaning Tunneling Microscopy (STM) 
\cite{Leibsle,Ellmer,Fishlock}
and Spot Profile Analysing Low Energy Electron Diffraction (SPA-LEED)
\cite{Croset} show mesoscopic morphologies 
different from droplet or stripe structures. 
For the N/Cu(001) case, the Nitrogen is chemically
adsorbed on a  $(001)$ Copper crystal surface, and agregates within 
square-shaped islands that may arrange either in 1-dimensional rafts
at low Nitrogen coverage or in a 2-dimensional array at intermediate 
coverages. 
The experimental works mentioned above \cite{Ellmer,Croset} motivated the present study.

Here we propose to describe the SO kinetics
on solid surface as a 2-dimensional spinodal decomposition.
In this standard theory, we include the stress energy due to the
underlying crystal and therefore
the surface elastic Green function is calculated. 
The 2D Cahn-Hilliard equation which drives the surface diffusion,
is integrated with computer means.
Our method
allows us to capture the SO kinetics together with
the elastic anisotropy due to the crystal symmetries. 
The latter feature is proved to 
play a role in the nanometric arrangements.
For simplicity, our study is focused on a (001) cubic crystal surface.

In the last decade, a similar theory has been developed under 
the name of ``phase field''
by Khachaturyan {\it et al.} for the kinetics of phase transition in alloys 
\cite{turyan83}. The phase field name
is also used
for models of solidification; see \cite{Misbah} for recent developments. 
We choose not to use this ambiguous terminology.
In the framework of the SO spinodal theory, the
implied coexistent phases
may represent either two types of crystal facet, 
noted A and B or a chemical adsorbed layer A over a clean crystal surface B. 
In the latter case, we neglect the possible layer  thickness.

Our results
exhibit a steady state at late stage of the separation process which 
shows different mesoscopic patterns
according to the coverage and to the elastic constants of the materials. 
Because of the crystal cubic symmetry, the nanometric 
morphologies of the final state may differ
from the one predicted in Refs. \cite{Marchenko81,Alerland88,Vanderbilt}.
At low coverage, rafts of either disk or square shaped  islands
appear while at intermediate coverage a branched stripe structure occurs.
The elastic constants of the material are shown to determine 
the preferential orientations of the rafts and of the stripes.
If the  surface square symmetry is broken, the
kinetics final state is similar
to those predicted in  Refs.\cite{Marchenko81,Alerland88}, i.e. , an assembly 
of regular spaced stripes.

\section{ Domain boundary energy}

The covering parameter $\theta$ describes 
the A-B coexistent phases at the solid surface.
These phases may differ by their composition or their geometry. 
Let say that $\theta=0$ for B and $\theta=1$ for A. 
If one assumes that the spatial surface variations of $\theta$ are smooth 
with respect to the atomic scale, 
a 2-dimensional coarse-graining procedure is thus relevant 
to represent the mesoscopic system state. 
If $d$ is the size of the elementary coarse-grained surfaces, we
introduce the mesoscopic quantity
$\theta(\mathbf{r})$ which is the $\theta$-average, 
performed over a whole elementary surface, centered at position $\mathbf{r}$. 
The  $\theta (\mathbf{r})$ variable is the mesoscopic local
coverage of the surface since it is the A quantity per unit area which is present
in the $\mathbf{r}$ vicinity. 
The $\theta $-average over the entire
surface  is written  $\theta_0$ and it is assumed to be conserved during the system time evolution.

An inhomogeneous mixing of A and B phases  involves an energy increasing
because of atomic bond breaking at domain boundaries. 
This energy together with the entropic term due to the surface inhomogeneity
are capted 
in the 2-dimentional free Ginzburg-Landau (GL) functional energy:
\begin{equation}
F_{chem}=F_0 . \int\int_S \{ \frac{\gamma}{2}  [(\frac{\partial \theta}{\partial x_1})^2 +(\frac{\partial \theta}{\partial x_2})^2]
+ {\hat f}(\theta) \}d \mathbf{r}\label{eq8}
\end{equation}
We introduce here the adimensional
free energy density $\hat f = 16 .\theta^2(1-\theta)^2$ which is a double well potential with minima
for $\theta=0$ and $\theta=1$. The hat notation points out the adimentional quantities.
The $\hat f$ form has no direct
influence on the mesoscopic structure providing GL functional is
invariant with respect to the (001) surface space group.
Let note $\hat g(\theta)$, the continuous Hamiltonian including both the 2D
gradient term and the free energy density $\hat f$.

The  $F_0$  and $\gamma$ scalars are
respectively the free energy density constant and 
the amplitude of the gradient term that both must be adjusted to
set the model domain boundary energy $I$
to a realistic value, i.e., around  $10$ meV/$\AA$ (see Ref. \cite{Alerland88}).
At phase equilibrium, the I quantity is given by  
$I = F_0 \int_B^A \ \hat g(\theta)  d\vec l$ 
where the integration is performed along a line path that goes from
inside a B phase domain to inside an A domain \cite{Lebowitz}. 
It is easy to see that the  previous path integral is 
overvalued by the product of the A-B interface width
times the $\hat f$ maximum value, i.e.,  $\hat f (\theta=0.5)=1$. 
Fixing $\gamma = 20\ d^2$, which insure the  $\theta({\mathbf r})$ space variation smoothness,
the A-B interface width is then
around  $5 \ .\ d$, at equilibrium. 
Therefore, to obtain $I \approx 10$ meV/$\AA$,
$ F_0 .d$ must be fixed to $2$ meV/$\AA$ (or $3.2\ 10^{-12}$ J/m).
As our investigations are focused over the nanometer scale, we choose to set
$d=1$ nm and thus $F_0 = 3.2$ mJ/m$^2$. 
To study larger space scales as it 
would  be suitable for Silicon which may exhibit a 100 nm
vicinal period \cite{Alerland88}, then it is sufficient to 
increase the $d$ parameter and to adjust subsequently 
$F_0$, eventually changing  the domain boundary energy $I$ if necessary.

For simplicity, we choose to neglect the possible
$\gamma$ variations with respect to the crystal surface direction.
Such a feature may be yielded from a crystal step anisotropy but here
our study is focused  on the elastic anisotropy due to the crystal symmetries.

\section{Surface Green function}
\label{Green}
An important stage of this work consists in the calculation of the  cubic
crystal surface elastic Green function. Let note 
 ${\bf P} (\mathbf{ r})$ a surface external force at position $\mathbf{ r}$.
In cartesian coordinates, the $(001)$ 
surface is defined by $(x_3=0)$ and ${\mathbf r}=(x_1,x_2)$.
The semi-infinite crystal occupies the half space $x_3 \ge 0$. 
The surface normal is the ${\bf n}$ unit vector.
The mechanic equilibrium
condition at the surface is given by:
\begin{equation}
\sigma_{i,j}( {\mathsf{r}},x_3=0). n_j=P_i({\mathsf{r}}) \label{eq1bis}
\end{equation}
where ${ n_j}$ is a ${\bf n}$ component
and the summation over subscript $j$ is implicit. 
The crystal bulk stress, $\sigma_{i,j}$($\mathsf{ r}$,$x_3$) 
is due to the crystal displacements 
and these quantities are related to each other by 
the Hooke law: $\sigma_{i,j}=\lambda_{i,j,k,l} \partial u_k / \partial x_l$. 
The forth order tensor
$\lambda_{i,j,k,l} $ gives the crystal elastic constants  and for
a cubic crystal symmetry,
this tensor is composed with three non zero coefficients \cite{Landau}, namely
$\lambda_{i,i,i,i}=C_{11}$, $\lambda_{i,i,j,j}=C_{12}$ and $\lambda_{i,j,i,j}=
\lambda_{i,j,j,i}=C_{44}$.
In case of copper, those coefficients are :
$C_{11}=1.683\ 10^{11}$ J/m$^3$, $C_{12}=1.221\ 10^{11}$ J/m$^3$ and 
$C_{44}=0.757\ 10^{11}$ J/m$^3$ (see Physics Handbooks).

At mechanic equilibrium, the bulk displacements fulfill the Lam\'e equation:
\begin{equation}
\lambda_{i,j,k,l} \frac{\partial^2 u_k}{\partial x_j \partial x_l}=0 
\label{eq2}
\end{equation} 
Except constants,
the displacement functions are fully determined by the set of equations  
(\ref{eq1bis},\ref{eq2}). As proposed in 
Ref.\cite{Villain} for the isotropic case,
these equations may be solved by writing the displacements as 2-dimensional Fourier transforms:
\begin{equation}
u_i({\mathsf r}, x_3)=\int\int \exp{(i {\bf Q}.{\mathsf{ r}})} . {\tilde u}_i({\bf Q}, x_3) d 
\mathsf{Q}\label{eq3}
\end{equation}
Note that the Fourier components ${\tilde u}_i({\bf Q}, x_3)$ depend on both the
wave vector ${\bf Q}=(q_1,q_2)$ and $x_3$.
Once the set of Eqs. (\ref{eq1bis},\ref{eq2}) is expressed using 
Eq.(\ref{eq3}), the subsequent differential equations involving ${\tilde u}_i$ and their derivatives 
with respect to $x_3$ turn out to be linear, so that ${\tilde u}_i({\bf Q}\ne 0, x_3)$  
have an exponential dependence on the bulk penetration length $x_3$:
\begin{equation}
{\tilde u}_j({\bf Q}\ne 0, x_3)=\sum_l \beta_{j,l} \exp{(-\alpha_l. x_3)}\label{eq4}
\end{equation}
where both $\beta_{j,l}$ and $\alpha_l$ depend on the wave vector ${\bf Q}$.
As the whole sample is at rest,
the ${\mathbf P}$ average
is assumed to be  zero which implies ${\tilde u}_j({\bf 0}, x_3)=0$. 
The Eq. (\ref{eq2}) is then reduced to a linear equation 
$M({\bf Q}, \alpha_l) (\beta_{j,l}) =0$ where $M({\bf Q}, \alpha_l)$ is the $3\times 3$ matrix:

\begin{eqnarray}  
M({\bf Q}, \alpha_l)=\left( \begin{array}{ccc}
C_{11} q_1^2 + C_{44} q_2^2 - C_{44} \alpha^2_l & (C_{12}+C_{44})q_1 q_2 & (C_{12} + C_{44}) i q_1 \alpha_l \\
(C_{12}+C_{44})q_1 q_2  & C_{11} q_2^2 + C_{44} q_1^2 - C_{44} \alpha^2_l  & (C_{12} + C_{44}) i q_2 \alpha_l \\
-(C_{12}+C_{44}) i q_1 \alpha_l  & -(C_{12}+C_{44}) i q_2 \alpha_l & C_{11} \alpha^2_l - C_{44} (q_1^2+q_2^2)\\
\end{array} \right) \label{stress-layer}
\end{eqnarray}
As usual, the non-trivial solutions
are such as $det\ (M({\bf Q}, \alpha_l))=0$. This yields a third degree polynomial equation 
for the $\alpha_l^2$ parameters that we solve numerically.
The only three $\alpha_l$ values with
positive real parts are physically acceptable.
At least one of the $\alpha_l$ is real, 
the two other roots may be either real or complex conjugated depending on 
the sign of the $\chi$ parameter:

\begin{equation}
\chi=C_{11}-C_{12}-2C_{44}\label{chi}
\end{equation}
This combination of the elastic constants is
related to the elastic anisotropy of the cubic crystal (see 
Ref. \cite{Landau}). When $\chi = 0$,
the crystal is isotropic and  the three 
$\alpha_j$ are degenerate.  This case was addressed in Refs. \cite{Landau,Villain}.
One retains as examples that the Copper and Gold $\chi$'s are negative 
$(\chi_{Cu}=-1.0,\ \chi_{Au}=-0.5 )$
and the Chromium and Niobium $\chi$'s are positive $(\chi_{Cr}=+1.8, \ \chi_{Nb}=+0.5)$.

To each $\alpha_j$ corresponds a unique set of three coefficients $(\beta_{j,l})$ that are
determined by inverting Eq. (\ref{eq1bis}).
Noting ${\tilde P}_j$ the Fourier transform of the force component $P_j$,
we write the Fourier counterpart of the  Eq. (\ref{eq1bis}):
\begin{eqnarray}
\sum_{l=(1,2,3)}  -\alpha_l  \beta_{l,1} +iq_1  \beta_{l,3} &=& {\tilde P}_1\nonumber\\
\sum_{l=(1,2,3)}  -\alpha_l  \beta_{l,2} + iq_2  \beta_{l,3} &=& {\tilde P}_2\nonumber\\
\sum_{l=(1,2,3)}  C_{12}(iq_1  \beta_{l,1} + iq_2  \beta_{l,2}) -C_{11} \alpha_l \beta_{l,3} &=& {\tilde P}_3
\nonumber\\
\label{FTsurfEq}
\end{eqnarray}

Combining the first and second rows
with the third rows of  the matrix $M({\bf Q}, \alpha_l)$ 
(see Eq. (\ref{stress-layer})), one easily gets both equations
\begin{eqnarray}
\beta_{l,1}=i q_1 . \beta_{l,3} \ \Gamma_{l,1}/\alpha_l\nonumber\\
\beta_{l,2}=i q_2 . \beta_{l,3} \ \Gamma_{l,2}/\alpha_l \nonumber\\
\label{Eqdrt}
\end{eqnarray}
where we write:
\begin{eqnarray}
\Gamma_{l,1}&=&\frac{(C_{11}-C_{12}-C_{44})\alpha_l^2-C_{44}(q_1^2+q_2^2)
}{(C_{11}-C_{12}-C_{44})q_1^2+C_{44}(q_2^2-\alpha_{l}^2)}\nonumber\\
\Gamma_{l,2}&=&\frac{(C_{11}-C_{12}-C_{44})\alpha_l^2-C_{44}(q_1^2+q_2^2)
}{(C_{11}-C_{12}-C_{44})q_2^2+C_{44}(q_1^2-\alpha_{l}^2)}\nonumber\\
\end{eqnarray}
provided the denominators of $\Gamma_{i,1}$ and $\Gamma_{i,2}$ are non zero.
If not, which occures for either $q_1=0$ or $q_2=0$ then
it is easy to show that $\beta_{l,3}=0$ and the calculation can be performed
as well. In addition let us note 
\begin{eqnarray}
\Gamma_{l,3}&=&-C_{11} \alpha_{l}-C_{12} (q_1^2 \Gamma_{l,1}+
q_2^2 \Gamma_{l,2})/\alpha_{l} \nonumber\\
\end{eqnarray}
We rewrite Eq. (\ref{FTsurfEq}) as a linear equation $ N_{j,l} \beta_{l,3}={\tilde P}_j$
where the $3\times 3$ matrix $N$ is defined as follows:
\begin{eqnarray}
N({\bf Q})=\left( \begin{array}{ccc}
iq_1 (1-\Gamma_{1,1})  & iq_1 (1-\Gamma_{2,1}) & iq_1 (1-\Gamma_{3,1}) \\
iq_2 (1-\Gamma_{1,2})  & iq_2 (1-\Gamma_{2,2}) & iq_2 (1-\Gamma_{3,2}) \\
\Gamma_{1,3}&\Gamma_{2,3} &\Gamma_{3,3}\\
\end{array} \right) 
\label{FTsurfEq2}
\end{eqnarray}
Provided $q_1 \ne q_2$, this matrix (\ref{FTsurfEq2}) can be inverted
which gives
the $\beta_{l,3}$'s as linear
functions of the ${P}_j({\mathbf r})$ Fourier transforms : $\beta_{l,3} = N^{-1}_{l,j} {\tilde P}_j$.
One must distinguish the case $q_1 = q_2$ for which details are not presented but are  simple to deal with.
Using Eqs. (\ref{Eqdrt}),
one gets the whole set of $\beta_{l,j}$'s and
thus ${\tilde u}_k ({\bf Q},x_3=0)= G_{k,j}({\bf Q}) {\tilde P}_j (\bf Q)$
where $ G_{k,j}$ is the surface elastic Green function
that we write as follows:
\begin{eqnarray}
G_{1,j}({\bf Q})&=& i q_1 \sum_l N^{-1}_{l,j} \ \Gamma_{l,1}/\alpha_l \nonumber\\
G_{2,j}({\bf Q})&=& i q_2 \sum_l N^{-1}_{l,j} \ \Gamma_{l,2}/\alpha_l \nonumber\\
G_{3,j}({\bf Q})&=& \sum_l N^{-1}_{l,j} \nonumber\\
\label{GreenFunction}
\end{eqnarray}
The latest function  varies as $1/| {\bf Q}|$.
The total elastic energy of the system is given at mechanical equilibrium according to Eqs. 
(\ref{eq1bis}, \ref{eq2}) by :
\begin{equation}
E_{el}= - 1/2 \int_{x_3=0} P_i .u_i dS   \label{eq6}
\end{equation}
that gives an analytical expression of $E_{el}$ in the Fourier space:
\begin{equation}
E_{el}= - 1/2 \int_{x_3=0} {\tilde P}_i^*  [G_{i,l} ] {\tilde P}_l  d{\bf Q}   \label{eq7}
\end{equation}

\noindent
The total energy is in fact
given by the sum $F=F_{chem}+E_{el}$ (Eqs. (\ref{eq8}) and (\ref{eq2})). 

The 
$\bf P$ force is induced by the phases
intrinsic stress misfit. Let note $\sigma_A^0$ and $\sigma_B^0$ the intrinsic
stress tensor of the A and B phases, respectively. 
At position $\bf r$,
the intrinsic stress is thus given by 
$\sigma^0({\bf r})=\sigma_B^0 + \delta\sigma^0 \theta({\bf r})$
where we introduce the tensor $\delta\sigma^0=\sigma_A^0-\sigma_B^0$.
The induced force is simply obtained by deriving $\sigma^0$ 
with respect to the surface coordinates, it gives:
\begin{equation}
P_i = \sum_{j=1,2} \delta \sigma^0_{ij} \frac{\partial \theta}{\partial x_j}
\end{equation}

Here we propose to distinguish whether 
there is a strong directional anisotropy 
such as for vicinal surface or
no broken surface symmetry. In the former case, if
a negligible stress variation is induced along a direction which
angle with respect to [100] is noted $\alpha$,   
the $\bf P$ vector must be zero if 
$\theta$ gradient points this direction.
The $\delta\sigma^0$ tensor coefficients are developed to satisfy
this condition and one may
write $\bf P$ as follows:

\begin{equation}
P_i = \Lambda_i [sin(\alpha)\frac{\partial \theta}{\partial x_1}
-cos(\alpha)\frac{\partial \theta}{\partial x_2}   ]
\end{equation}
where we introduce three constants $\{\Lambda_i\}$.
We restrict our study to a uniaxial stress with only one non zero  $\Lambda_i$
which is  $\Lambda_3=\Lambda$.
According to our tests,  non zero 
$\Lambda_1$ and $\Lambda_2$ involve no qualitative change in our results.

If the surface square symmetry is preserved,
the $\delta\sigma^0$ tensor is invariant under the 
square symmetry group operators.
So
$\delta\sigma^0_{12}=\delta\sigma^0_{21}$, 
$\delta\sigma^0_{31}=\delta\sigma^0_{32}$
and $\delta\sigma^0_{11}=\delta\sigma^0_{22}$.
Again we introduce the $\Lambda$ parameter setting that
$\delta\sigma^0_{11}=\delta\sigma^0_{22}=\Lambda$.
The surface is assumed to be a perfect plane 
so the $P_3$ component is 
negligible, i.e., $\delta\sigma^0_{31}=\delta\sigma^0_{32}=0$. 
Non zero coefficients
$\mu=\delta\sigma^0_{12}=\delta\sigma^0_{21}$
implies a shear that may be induced
by the internal structures of the coexistent phases.
To a first step, we choose to ignore such a possibility, i.e.,
$\mu << \Lambda$.
Thus the force $\bf P$ is simply proportional to the $\theta$ gradient.

The parameter  $\Lambda$
is a constant that fixes the amplitude of the surface 
force and it may vary from an experimental case to another. 
In our model,
for each different set of crystal elastic coefficients,  
$\Lambda$  is adjusted, i.e.,
our numerical calculations are performed with
$\Lambda = 40\ mJ/m^2$ (or $\Lambda =0.25 eV/\AA^2$ ) for the Copper, 
Gold and Niobium crystal cases 
and $\Lambda = 57\ mJ/m^2$ (or $\Lambda =0.35 eV/\AA^2$ ) for the  Chromium case.

\section{Kinetics of phase separation}

To modelize the phase separation kinetics, we use a standard
spinodal decomposition theory
which is precisely described in Ref.\cite{Lebowitz} for the 3-dimensional case. 
Extension to the 2-dimensional systems is straighfoward.
The time evolution of the conserved quantity $\theta( \mathsf r)$
is then driven by the Cahn-Hilliard equation (also known as model B):
\begin{equation}
\frac{\partial \theta ({\mathsf r} ,t)}{\partial t}=
M \bigtriangleup \frac{\delta F }{\delta \theta({\mathsf r} ,t)} 
+ \epsilon({\mathsf r} ,t) \label{KinC}
\end{equation}
where $\epsilon$ is a Langevin stochastic term which simulates thermal fluctuations
as proposed by H.E. Cook in \cite{Cook}:
\begin{eqnarray}
<\epsilon ( {\mathsf{ r}},t)>&=&0\\
<\epsilon( {\mathsf{ r}},t) \epsilon({\mathsf{ r}}' ,t') >&=&-2 k_B T M
\Delta \delta ( {\mathsf{ r}} -{\mathsf{ r}}')\delta (t -t') \label{noiseCor}
\end{eqnarray}

In the previous Eq. \ref{KinC}, $M$ is a mobility constant 
which value may be related to
the Fick diffusion constant $D_{Fick}$. As known, for a given chemical species
the parameter  $D_{Fick}$
strongly depends on both temperature and coverage (see Ref. \cite{Zangwill}). 
For simplicity, 
we assum here that  $D_{Fick} \approx 10^{-6} cm^2/s$  (see \cite{Zangwill}).
Neglecting the elastic term $E_{el}$, a linear expansion of Eq. \ref{KinC} around a 
uniform coverage $\theta ({\mathsf r}) = 0.5 $ shows that the 
Eq. \ref{KinC} is equivalent to a diffusion equation for 
short wavelength perturbations (see \cite{Lebowitz}). Then, we establish
$ M . F_0. \frac{d^2 {\hat f}}{d\theta^2}  =D_{Fick}$  
which fixes approximately the mobility since
$\frac{d^2 {\hat f}}{d\theta^2} (\theta_0=0.5)= 16$, in the present case. 
As both $D_{Fick}$ and $\frac{d^2 {\hat f}}{d\theta^2}$ strongly decrease 
with a decreasing $\theta_0$, the mobility variations with respect to the average coverage 
can not be estimated in the general context of our work. For simplicity, we choose
to keep it constant, i.e., it is equal to the mobility estimated at half coverage. 

Regarding the kinetics of the coarse-grained surface, it is simply derived
from the space and time discretization of Eq. \ref{KinC}. 
The  previous linear development of Eq. \ref{KinC} is still valid
and it comes that the diffusion constant $D_{Fick}$
is rescaled as $D_{Fick} /d^2 $ due to 
the discretized Laplacian operator. It corresponds to the intuitive idea according to which
the larger the coarse-graining is, i.e., the larger is $d$, the slower each unit cell of the coarse-graining
evolves and thus the slower is the global kinetics.
One guesses that the larger is a B domain, the longer is the time needed to 
transform its whole area in a A phase region.
Thus we are able to propose a rough estimation of the surface kinetics
time scale. Our computer calculations
work with a time unit which is equivalent to 
$\delta t = d^2 / D_{Fick} \approx  10^{-8} s $. 
To perform the time integration of Eq. \ref{KinC}, we use a time increment of 
$(10^{-3} \times \delta t)$.

The Eq. (\ref{KinC}) is integrated starting from 
a uniform initial state for which $\theta ({\mathsf{ r}})= \theta_0$.
Spontaneous formation of domains is observed for a coverage $\theta_0$ 
lying between $0.25< \theta_0 < 0.75$. This range is imposed by the 
classical spinodal region defined by  $\partial^2 {\hat f}/\partial^2 \theta<0$ 
(see \cite{Lebowitz}).
These limits are indeed slightly modified in presence of elastic interactions.
The temperature dependency may be activated
in the functional $F_{chem}$, in 
the diffusion coefficient $D_{Fick}$, and in the amplitude of the Langevin noise.
The external parameters are the material elastic constants that fix $\chi$,
the average coverage $\theta_0$, the surface stress  $\Lambda$ 
and the $\alpha$ angle for both isotropic and anisotropic stress cases.


Our numerical findings  confirm the analytical results of 
Refs. \cite{Marchenko81,Alerland88,Vanderbilt}: the
phase separation is frustrated by long range interaction
due to the crystal surface stress, i.e.,
the kinetics yields a final steady state.
As model input parameters, one may retain that
the domain boundary energy is of order $I=10 meV/{\AA}$ 
and the stress amplitude is either
$\Lambda = 40\ mJ/m^2$ (or $\Lambda =0.25 eV/\AA^2$ ) for the Copper, 
Gold and Niobium crystal cases or
$\Lambda = 57\ mJ/m^2$ (or $\Lambda =0.35 eV/\AA^2$ ) for the  Chromium case.

In the case of a broken surface symmetry with
uniaxial stress,   
the kinetics yields non branched stripes structures with spacial
periodicity along a given direction 
as shown on Fig.\ref{fig1} (for $\alpha=\pi /4$). 
The selected period strongly decreases when $\Lambda$ increases 
but the formula proposed in Refs. \cite{Marchenko81,Alerland88} which 
gives the period versus the model parameters,
has not been confirmed by our computations because it required many
long time computer runs. It is the purpose of a forthcoming paper. 
For a coverage $\theta_0=0.3$
(see second row in Fig.\ref{fig1}), the time needed to reach
the final stripe structure is longer than for $\theta_0=0.5$. 
As soon as the nucleation occures, it appears some distorted stripes 
length of which
increases with coverage. 
During the coalescence regime, at  $\theta_0=0.5$,
the stripes branch to each other before their connections
move along the stripe sides such that it eliminates the domain
which separates the primary stripe couple. The mecanism is very different
at low coverage (see second row
of Fig.\ref{fig1}), since the short domains move
to align themself with longer stripes.

For a preserved surface symmetry, the kinetics
computer calculations results are presented on Figs.\ref{fig2} and \ref{fig4}.
In Fig.\ref{fig2}, 
the time evolution of a half covered surface with and without induced stress is shown. 
One notes that for a negligible elastic interaction, i.e., $\Lambda=0$,
the domains never stop growing because of the Ostwald's ripening \cite{Lebowitz,LSW}. 
We had to stop the simulation when the size of the domains is of the order of
the whole sample.
For $\Lambda \neq 0$, the kinetics 
is radically  different since the system is driven 
to a ``self-organized'' final state. 
Beyong an evolving stage
which duration decreases
as $\Lambda$ increases, the surface reaches a steady structure 
as shown by the two last pictures of the second and the third rows in Fig. \ref{fig2}.
The specific size and the distance between the neighboring domains
are determined
by the value of $\Lambda$. The larger is $\Lambda$, the smaller are the domain sizes.
According to our calculations, 
the final state 
is composed of branched stripes with tips
as shown by
the six last pictures in Fig.\ref{fig2}. 
As in Fig.\ref{fig1}, the SO appears as soon as the nucleation occures. At this stage,
the domains already arrange along preferential orientations. The structure evolves
such that the thinest branches disappear to the benefit of the thick stripes by 
first consuming  the tips.
The stripe orientations depend clearly on the
sign of $\chi$: for $\chi > 0$ as for Chromium, 
stripes are along either $[110]$ or $[1 \bar{1} 0]$
while  for $\chi < 0 $ as for Copper, stripes are along either $[100]$ or $[010]$.
According to the few
experimental results about SO of 
chemisorbed mono-layer, 
the adsorption of Nitrogen on a (001) Chromium surface
has been  analysed with STM  by M. Pinczolits {\it et al.} in Ref. \cite{pinczolits}. 
Infortunately, none of their results allows to conclude about the possible 
arrangement of Nitrogen atoms along [110] or $[1 \bar{1} 0]$ directions. 
The experimental results obtained by \cite{Leibsle,Ellmer} about (001) Copper surface are 
in a good agreement with our calculations.

In Fig.\ref{fig3}, the plot of the ratio  $\delta F/(F_0.S)$ where
$S$ is the total surface size 
shows how the relative free energy $\Delta F$ evolves with time. This quantity
$\delta F$ is the total free energy measured with respect to the free 
energy of a surface with only two compact domains of A and B phases. 
The stressed surface decomposition kinetics
generates a steady state with
lower energy because of a negative elastic energy $E_{el}$. 
After a nucleation regime where the total energy decreases 
strongly as time increases, the energy reaches an  asymptotic value.
The asymptotic state free energy  clearly depends on the coverage $\theta_0$
and the difference $\delta F(\theta_0=0.5)- \delta F(\theta_0=0.25)$
increases in absolute value with $\Lambda$. 
The same plot with no surface stress 
shows that the surface evolves in order
to reduce its positive domain boundary  energy.
In such a case, the  asymptotic energy difference of two
surfaces with different coverages, i.e.,  
$\delta F(\theta_0=0.5)-\delta F(\theta_0=0.25)$  becomes negligible.

At low coverage, as it was predicted
in Ref.\cite{Vanderbilt},
an assembly of dots appear in the final stage. 
As for stripes, 
the arrangement of the dots also depends on the sign of $\chi$. 
The dots are arranged in an assembly of 1-dimensional rafts: the raft orientation
is either along $[110]$ or $[1 \bar{1}0]$ for $\chi > 0$  (first row of Fig.\ref{fig4}),
and  either along $[100]$ or $[010]$ for $\chi <0$ (second row of Fig.\ref{fig4}).
In Fig.\ref{fig4},
the dots may be square shaped
because of the relaxation of  their elastic energy and the underlaying crystal symmetry.  
Comparison of
the total free energy of a single  A phase dot  
on a B surface, imposing different shapes to this dot,
shows that the square shape is optimal,
provided the area covered by
the dot is larger than a critical size.
For smaller sizes, the droplets are disk shaped.
This critical size  depends on both the elastic constants and $\Lambda$.
As a consequence, in the first stage of the growth, when the domains have
small area, all of them are disk shaped. 
In the case of Copper, before the surface reaches its steady state,
most of the islands  undergo a shape tranformation passing from a disk shape to a square shape 
(see the first row of Fig.\ref{fig4}). 
In such a case, the domain size 
is clearly larger
than the critical shape transformation size. In the left picture of the first row of Fig.\ref{fig4}, 
performed with Chromium elastic coefficients, 
it is the opposite case, since the disk shape domains never transform. In the right
picture of the same row, 
increasing the surface coverage $\theta_0$ yields larger domains. Some 
of them overpasses the critical size to becomes square with $[110]$ and $[1 \bar{1}0]$ 
oriented sides.

For a coverage lying between $\theta_0 = 0.25$ and half coverage, we find that
the nanostructure of the final stage
is composed of a mixing between short length stripes and dots (see Fig. \ref{fig4}). 
So, we deduce that
the surface SO  shows a crossover with respect to the coverage $\theta_0$, i.e.,
going from droplet structure at low coverage  to a stripe structure at half
coverage, passing through mixed structures.

The Fig. \ref{fig5} shows the displacement field involved by the presence of
the adsorbate. Nearby the surface, vortices are induced
inside the bulk and extend over a depth of few ten nanometers.
The maximum amplitude of the displacements is of few $0.1\ \AA$.
With both a molecular dynamics study and a linear elasticity analysis
of  the specific N/Cu (001) case,
those structures are also found by B. Croset {\it et al.}  \cite{Prevost} 
who first pointed them to us. Moreother
they showed that the bulk displacements due 
to vortices contribute to X-ray diffraction and they deduce a surface force
amplitude measurement: $2.2 \ 10^{-9}$ N.at$^-1$ (or $0.3$ eV/$\AA^2$). In our model,
the $\Lambda$ amplitude has same order of magnitude.

For (001) Copper surface case, an induced shear is tested, i.e., a
non zero  $\mu=\delta \sigma^0_{12}=\delta \sigma^0_{21}$ (defined above). 
Both the island shape and the arrangement directions are tilted with an angle that 
depends on the ratio $\mu/\Lambda$ (see Fig. \ref{fig6}).

The previous computations are performed with no thermal fluctuations.
Putting on the Langevin noise in Eq. (\ref{KinC}), 
with an amplitude which is related to a temperature  by Eq. (\ref{noiseCor}),
we performed the same simulations and obtained similar final states with
the same structures. This
demonstrates the stability of both the kinetics and the final state.
At half coverage and for the Copper elastic coefficients,
the amplitude of the thermal noise may be increased up to 
the melting point of the Copper metal. 
Our method does not simulate the desorption process which could
be activated at lower temperature than the crystal melting point.
Moreother the  white noise amplitude is calculated to satisfy 
the fluctuation dissipation theorem which is strictly valid  around
the thermodynamical equilibrium. Nevertheless the use of the Langevin noise
allows to estimate qualitatively the  surface structure ability to stand up to
thermal fluctuations. Here
the nanometric self-organised surface structure are proved to be very strong.

Similar results have been obtained with elastic constants of Gold ($\chi <0$ )
and with Niobium elastic constants ($\chi >0$) 
with a smaller amplitude $\Lambda$, indeed (see above for details).

\section{Conclusion and Perspectives}

In the present paper, we investigated  the patterns that 
are yielded from the $(001)$ cubic crystal surface SO. 
If  the surface square symmetry is preserved,
the nanometric morphologies are proved to depend on both the 
coverage and the elastic coefficients
of the crystal. One-dimentional rafts of 
dots (with disk shape or square shape) occur
at low coverage and branched stripes appear at half coverage. 
Rafts and stripes are aligned with
either [100] or [010] for a negative anisotropy factor and with either 
[110] or [1$\overline{1}$0] for a
positive  anisotropy factor.
Our computations have been performed with different cubic metal elastic coefficients, e.g.,
Gold, Copper, Chromium and Niobium.
Obviously, it is possible to extend our calculations
to the surfaces of either alloys or ionic crystal with cubic symmetry.
In case of metal,
this study should be extended to other
surface symmetries which 
exibits interesting properties such
as the (111) Gold surface 
e.g., the Gold Herringbone reconstruction (see, e.g., Refs. \cite{Au111,narasimhan}), and
as the (110) Copper surface, e.g.,
the Oxyde adsorption (see Ref. \cite{Cu110-O}). 
Then,  in the framework of the spinodal theory for SO,
only the elastic Green function should be modified.

In some specific cases, the crystal  surface  SO 
displays experimental features
that are not included in the present model.
The STM analysis of N/Cu(001) performed 
by Ellmer {\it et al.} \cite{Ellmer} 
demonstrates that the surface morphologies are not  equivalent for
coverage $\theta_0 = (0.5+x)$ and for $\theta_0 = (0.5-x)$,  
as one would expect from the spinodal decomposition theory we proposed.
In addition, 
attractive interactions between Nitrogen islands 
seems to be dominant at very low coverage. 
We also note a remarkable result of T.W. Fishlock {\it et al.} \cite{Fishlock} on Br/Cu(001)
where STM analysis shows bromine islands organized in a chessboard at half coverage
while we would expect branched stripes. We believe that
our simple method can be improved in order to investigate those specific cases even if
it is clear that a continuous approach is limited to the mesoscopic scale
analysis, e.g.,
it seems  hopeless to include certain atomic scale features as 
surface dislocations, or ad-atom missing row.

In the present work, the
chemisorbed layers thickness is neglected.
As shown in Ref. \cite{Tersoff} for the Silicon quantum dots,
this thickness plays a role in the arrangement of dots but 
for the moment no idea emerges to tackle this problem on the basis of 
our approach.

For simplicity, we ignored the possible
internal structures of each phases. If not 
it  may involve both anisotropy 
and shear that are expected to modify 
the mesoscopic patterns. Moreother,
some translational and orientational variants may be yielded
from the different way the ad-atoms 
arrange on the underlying surface crystal lattice.
Those variants can be treated in the present model
by developing the free energy density 
with respect to  the so called long range order parameters.
This extension is  
well known within the phase field method for alloy physics \cite{turyan83}.
As a consequence of the coexistent variants, some anti-phase boundaries
should appear between domains and play a role in the pattern growth.

In summary, 
the model we present here is a promising start 
to describe and also to predict the nanostructures induced by self-organization
of a crystal surface.

\newpage

\centerline{Figure captions}

{\label{fig1} Fig. 1 (A-F): Phase separation kinetics 
of a $( 0 0 1)$  cubic crystal surface ($\chi=\chi_{Au}\approx -0.5$)
with a stress directional anisotropy 
$\alpha= \pi/4$ (defined in the text). Pictures occurence time is indicated
at the low left hand corners: 
time unit is around $10$ ns. Kinetics is started at initial time $t=0$
with a uniform surface state.
The direction $[0 1 0]$ is indicated. }\\

{\label{fig2} Fig. 2 (A-I): Phase separation kinetics 
on a $(0 0 1)$  cubic crystal surface with 
negligible stress, i.e., $\Lambda=0$ (defined in the text)  first row, 
and with $\Lambda =40\ mJ/m^2 $  $\chi=\chi_{Cu}=-1$ , second row 
and $\Lambda = 57 \ mJ/m^2$  and $\chi=\chi_{Cr}=1$ third rows.}\\

{\label{fig3} Fig. 3: Time evolution of the total free energy $\delta F$
for negligible stress, i.e.,
$\Lambda=0$, and for a stressed surface, i.e., $\Lambda =40\ mJ/m^2 $
for different coverages: $\theta_0=0.25$ (dashed lines)
and $\theta_0=0.5$ (solid lines).
Time unit is $10$ ns and the free energy is divided by the constant 
$F_0.S$ (defined in the text).}\\

{\label{fig4} Fig. 4 (A-B) and 5 (A-B): First row: steady states of a (001) cubic crystal surface with  $\Lambda =40\ mJ/m^2 $,
$\chi=\chi_{Cu}=-1$,  $\theta_0 =0.21$ (on the left) and
$\theta_0 =0.35$ (on the right).\\
Second row: steady state of a (001) cubic crystal surface with  $\Lambda = 57 \ mJ/m^2$,
$\chi=\chi_{Cr}=1$ , $\theta_0 =0.21$ (on the left) and $\theta_0 =0.35$ (on the right).}\\

{\label{fig5} Fig. 6 : Profile of the displacement field inside the bulk, for $\Lambda =40\ mJ/m^2 $, 
$\chi=\chi_{Cu}=-1$ and $\theta_0=0.21$.
The arrows lenght is proportional to the
displacement (multiplied by one hundred). A distance $d=1 nm$
separates neighboring sites of the coarse-grained lattice.
The directions  $[ 0 1 0]$ and $[ 0 0 1 ]$ are indicated.}

{\label{fig6} Fig. 7: Steady state of a (001) cubic crystal surface with  $\Lambda =40\ mJ/m^2 $,
$\chi=\chi_{Cu}=-1$ , $\theta_0 =0.3$ and a shear component 
$\delta \sigma^0_{12}=\delta \sigma^0_{12}=\Lambda/2$.}\\

\end{document}